\documentclass[a4paper]{jpconf}
\usepackage{amssymb}
\usepackage{amsthm}
\usepackage{latexsym}
\newtheorem{theorem}{Theorem}
\def\be{\begin{equation}}
\def\ee{\end{equation}}
\def\bq{\begin{eqnarray}}
\def\eq{\end{eqnarray}}
\def\beq{\begin{eqnarray*}}
\def\eeq{\end{eqnarray*}}

\begin{document}
\title{\textbf{Bel-Robinson energy and the nature of
singularities in isotropic cosmologies}}
\author{\textbf{Ifigeneia Klaoudatou$\dagger$, Spiros Cotsakis$\ddagger$}}
\date{}
\address{Research Group of Mathematical
Physics and Cosmology, Department of Information and Communication
Systems Engineering, University of the Aegean, Karlovassi 83 200,
Samos, Greece} \ead{iklaoud@aegean.gr$\dagger$}
\ead{skot@aegean.gr$\ddagger$}
%\vspace{1cm}
%\numberwithin{equation}{section}
%%%%%%%%%%%%%%%%%%%%%%%%%%%%%%%%%%%%%%%%%%%%%%%%%%%%%%%
\begin{abstract}
We review our recent work on the classification of finite time
singularities that arise in isotropic universes. This scheme is
based on the exploitation of the  Bel Robinson energy in a
cosmological setting. We comment on the relation between geodesic
completeness and the Bel Robinson energy  and  present evidence
that relates the divergence of the latter to the existence of
closed trapped surfaces.
\end{abstract}
%%%%%%%%%%%%%%%%%%%%%%%%%%%%%%%%%%%%%%%%%%%%%%%%%%%%%%%

\section{Introduction}

%%%%%%%%%%%%%%%%%%%%%%%%%%%%%%%%%%%%%%%%%%%%%%%%%%%%%%%
In \cite{ck05} we provided necessary conditions for the appearance
of finite time singularities in isotropic universes based entirely
on the behaviour of the Hubble parameter. These conditions
provided us with a new classification of such singularities, a
scheme which was further refined and expanded in \cite{ck06} to
include all known types of singularities and predict new ones.
This second approach exploited the Bel Robinson energy, a kind of
energy of the gravitational field projected in a sense to a slice
in spacetime. Combining the quantities that constitute the Bel
Robinson energy with the pressure and density of the matter fields
through the field equations, we were thus led to a clear picture
of how the matter fields influence the nature of the
singularities. A complete classification was then obtained with
the use of the Hubble parameter and the Bel Robinson energy.

In the next Section, we introduce the Bel-Robinson energy and show
that  a closed Robertson-Walker (in brief \textsc{RW}) universe is
geodesically complete if this energy is bounded. We present a
summary of our classification scheme in Section 3, and in Section
4 we give necessary and sufficient conditions for the character of
the resulting singularities. In the last Section, we discuss the
existence of closed trapped surfaces in terms of the Bel-Robinson
energy.

%%%%%%%%%%%%%%%%%%%%%%%%%%%%%%%%%%%%%%%%%%%%%%%%%%%%%%%%%%%%%%%%

\section{Bel Robinson energy and completeness}

%%%%%%%%%%%%%%%%%%%%%%%%%%%%%%%%%%%%%%%%%%%%%%%%%%%%%%%

Consider a sliced spacetime with metric
\begin{equation}
^{(n+1)}g\equiv -N^{2}(\theta ^{0})^{2}+g_{ij}\;\theta ^{i}\theta
^{j},\quad \theta ^{0}=dt,\quad \theta ^{i}\equiv dx^{i}+\beta
^{i}dt,  \label{2.1}
\end{equation}
where $N=N(t,x^{i})$ is the lapse function and
$\beta^{i}(t,x^{j})$ is the shift function, and the $2$-covariant
spatial electric and magnetic tensors \beq E_{ij}&=&R^{0}_{i0j},\\
D_{ij}&=&\frac{1}{4}\eta_{ihk}\eta_{jlm}R^{hklm},\\
H_{ij}&=&\frac{1}{2}N^{-1}\eta_{ihk}R^{hk}_{0j},\\
B_{ji}&=&\frac{1}{2}N^{-1}\eta_{ihk}R^{hk}_{0j}, \eeq where
$\eta_{ijk}$ is the volume element of the space metric $\bar g$.
These four time-dependent space tensors comprise what is called a
\emph{Bianchi field}, $(E,H,D,B)$, a very important frame field
used to prove global in time results, cf. \cite{ycb02}.

The \emph{Bel-Robinson energy at time $t$} is given as the space
integral \be
\mathcal{B}(t)=\frac{1}{2}\int_{\mathcal{M}_t}\left(|E|^{2}+|D|^{2}+
|B|^{2}+|H|^{2}\right)d\mu_{\bar{g}_t}, \ee where by
$|X|^{2}=g^{ij}g^{kl}X_{ik}X_{jl}$ we denote the spatial norm of
the $2$-covariant tensor $X$. In the following, we focus on a
$\textsc{RW}$ geometry filled with various forms of matter and
with a metric given by \be ds^2=-dt^2+a^2(t )d\sigma ^2, \ee where
$d\sigma ^2$ denotes the usual time-independent metric on the
3-slices of constant curvature $k$. For this spacetime,  the norms
of the magnetic parts $|H|, |B|$ are identically zero while $|E|$
and $|D|$, the norms of the electric parts, reduce to \be
|E|^{2}=3\left(\ddot{a}/{a}\right)^{2} \quad\textrm{and}\quad
|D|^{2}=3\left(\left(\dot{a}/{a}\right)^{2}+k/{a^{2}}\right)^{2}.
\ee Therefore the Bel-Robinson energy becomes \be
\mathcal{B}(t)=\frac{C}{2}\left(|E|^{2}+|D|^{2}\right), \ee where
$C$ is the constant.

It is not difficult to show that in a closed \textsc{RW} universe
such that $|D|$ is bounded above, $H$ must be bounded above and
the scale factor bounded below. Therefore $H$ must be integrable
and the spatial metric bounded below, that is such a universe is
regularly hyperbolic. This in turn means that all hypotheses of
the completeness theorem proved in \cite{chc02} are satisfied and
therefore such a universe is $g-$complete. It is also
straightforward to see that the null energy condition is
equivalent to the inequality $|E|\leq |D|$, hence completeness is
then accompanied with $|E|$ being bounded above. (For a flat
spatial metric we would need to impose the regular hyperbolicity
hypothesis in order to conclude completeness since the latter is
independent from the boundedness of $|D|$.)
%%%%%%%%%%%%%%%%%%%%%%%%%%%%%%%%%%%%%%%%%%%%%

\section{Classification of singularities}

%%%%%%%%%%%%%%%%%%%%%%%%%%%%%%%%%%%%%%%%%%%%

We are interested in finding the possible types of singularities
that can arise in the evolution of a $\textsc{RW}$ geometry. These
types result from the possible behaviours exhibited by the
different triplets consisting of the scale factor $a$, the Hubble
expansion rate $H$ and the Bel Robinson energy $\mathcal{B}$.
Assuming that the model has a finite time singularity at $t=t_s$,
the possible behaviours of the functions in the triplet
$\left(H,a,(|E|,|D|)\right)$ are classified as follows:
\begin{description}
\item [$S_{1}$] $H$ non-integrable on $[t_{1},t]$ for every $t>t_{1}$

\item [$S_{2}$] $H\rightarrow\infty$  at $t_{s}>t_{1}$

\item [$S_{3}$] $H$ otherwise pathological
\end{description}

\begin{description}
\item [$N_{1}$] $a\rightarrow 0$

\item [$N_{2}$] $a\rightarrow a_{s}\neq 0$

\item [$N_{3}$] $a\rightarrow \infty$
\end{description}

\begin{description}
\item [$B_{1}$] $|E|\rightarrow\infty,\, |D|\rightarrow \infty$

\item [$B_{2}$] $|E|<\infty,\, |D|\rightarrow \infty $

\item [$B_{3}$] $|E|\rightarrow\infty,\, |D|< \infty $

\item [$B_{4}$] $|E|<\infty,\, |D|< \infty $.
\end{description}
The nature of a prescribed  singularity is thus characterized
completely by specifying the components in  triplets of the form
\[(S_{i},N_{j},B_{l}),\] with the indices $i,j,l$ taking their
respective values as above.

For fluid-filled \textsc{RW} models the various behaviours of the
Bel-Robinson energy density can be written equivalently in terms
of  the density and pressure of the cosmological fluid:
\begin{description}
\item [$B_{1}$] $\Leftrightarrow$ $\mu\rightarrow \infty$ and
$|\mu+3p|\rightarrow\infty$

\item [$B_{2}$] $\Leftrightarrow$ $\mu\rightarrow \infty$ and
$|\mu+3p|<\infty$

\item [$B_{3}$] $\Leftrightarrow$ $\mu<\infty$ and
$|\mu+3p|\rightarrow\infty$ $\Leftrightarrow$ $\mu<\infty$ and
$|p|\rightarrow\infty$

\item [$B_{4}$] $\Leftrightarrow$ $\mu< \infty$ and $|\mu+3p|<\infty$
$\Leftrightarrow$ $\mu<\infty$ and $|p|<\infty$.
\end{description}

Following \cite{ck06}, we may gain further insight into the nature
of any of the singularities expounded above by  studying  the
asymptotic behaviour of the three functions that determine the
character of the singularity. This is done by finding the relative
strength, i.e., the limit as $t\rightarrow t_{s}$ of the relative
rate of decay, for each pair of these functions. As an  example
compare the standard big bang singularities in the  dust and
radiation models in general relativity; even though both of
singularities are of the type $(S_{1},N_{1},B_{1})$, they are
really different since the radiation one has relative strength
$a<<H<<(|E|\leftrightarrow |D|)$ (where `$<<$',
`$\leftrightarrow$', `$\sim$' mean that the relative limit is $0$,
$1$ or $c\neq 0,1$, with $c$ a constant, respectively), whereas
that in the  dust models is characterized by the asymptotic
behaviour $a<<H<<(|E| \sim |D|)$.

%%%%%%%%%%%%%%%%%%%%%%%%%%%%%%%%%%%%%%%%%%%%%%%%%%%%%%%%%%%%%%%%%%%%%%%%%5

\section{Necessary and sufficient conditions for an $(S_{i},N_{j},B_{l})$
type of singularity}

%%%%%%%%%%%%%%%%%%%%%%%%%%%%%%%%%%%%%%%%%%%%%%%%%%%%%%%%%%%%%%%%%%%%%%%%%%

Flat \textsc{RW} universes are clearly the simplest to analyze but
in the long run
 one is interested to see whether the behaviours known to hold in the flat
case continue to remain valid in more generic curved models. In
this Section, we give necessary and sufficient conditions for the
 singularities  in  given flat \textsc{RW}
models to continue to hold in more general universes in the sense
of either  having nonzero values of $k$ or described by solutions
in which some or all of the arbitrary constants remain arbitrary
(particular or general solutions, cf. \cite{cb06} for this
terminology). Usually the curvature term in a typical
Friedman-type equation turns out to be \emph{subdominant} compared
to the density term or, in any case, does not alter the $H$
behaviour. Some examples of what can happen in general
relativistic models are shown in the following theorems.

\begin{theorem} \label{2}
Necessary and sufficient conditions for an $(S_{2},N_{3},B_{1})$
singularity occurring at the finite future time $t_{s}$ in an
isotropic universe filled with a fluid with  equation of state
$p=w\mu$, are that $w<-1$ and $|p|\rightarrow\infty$ at $t_{s}$.
\end{theorem}

\begin{theorem} \label{3}
A necessary and sufficient condition for an $(S_{2},N_{2},B_{1})$
singularity at $t_{s}$ in an isotropic universe filled with a
fluid with equation of state $p+\mu=-B\mu^{\beta}$, $\beta>1$, is
that $\mu\rightarrow\infty$ at $t_{s}$.
\end{theorem}

\begin{theorem} \label{4}
A necessary and sufficient condition for an $(S_{3},N_{2},B_{3})$
singularity at $t_{s}$ in an isotropic universe filled with a
fluid with equation of state $p+\mu=-C(\mu_{0}-\mu)^{-\gamma}$,
$\gamma>0$, is that $\mu\rightarrow\mu_{0}$ at $t_{s}$.
\end{theorem}

\begin{theorem} \label{5}
A necessary and sufficient condition for an $(S_{1},N_{1},B_{1})$
singularity at $t_{1}$ in an open or flat isotropic universe
filled with a fluid with equation of state
$p+\mu=\gamma\mu^{\lambda}$, $\gamma>0$ and $\lambda<1$, is that
$\mu\rightarrow\infty$ at $t_{1}$.
\end{theorem}

\begin{theorem} \label{6}
A necessary and sufficient condition for an $(S_{1},N_{1},B_{1})$
singularity at $t_{1}$ in an isotropic universe with a massless
scalar field is that $\dot{\phi}\rightarrow\infty$ at $t_{1}$.
\end{theorem}

We can easily find examples of the behaviours described in the
above theorems. The phantom cosmologies of  \cite{gonzales} are
described  by Theorem (\ref{2}), the various solutions of
\cite{noj1} can be all accommodated by theorems  (\ref{3}) and
(\ref{4}) respectively. The graduated inflationary models
constructed by Barrow in \cite{ba} when $\lambda=3/4$   belong to
the situation of Theorem (\ref{5}) whereas the singularities of
massless scalar field model of Ref. \cite{fo} are described by
Theorem (\ref{6}). We shall meet the latter two models again in
the next Section. The proofs of theorems (\ref{2})-(\ref{6}) and
the details of the dark energy examples can be found in
\cite{ck06}, our basic reference for all these results.

One of the virtues of our classification scheme is that it can
pick up even very mild singularities. The mildest in a sense type
of singularity known to the authors was discovered in \cite{noj1}.
This is a general relativistic, flat \textsc{RW} universe filled
with a fluid satisfying the equation of state
$$p+\mu=-\frac{AB\mu^{2\beta-1}}{A\mu^{\beta-1}+B},\quad
0<\beta<1/2,$$ and for $\beta=1/5$ it admits the exact solution
$a=a_{0} e^{\tau^{8/3}}$. Then $H=(8/3)\tau^{5/3}$,
$\dot{H}=(40/9)\tau^{2/3}$ and $\ddot{H}=(80/27)\tau^{-1/3}$. Thus
as $\tau\rightarrow 0$, $a$, $\dot{a}$, $\ddot{a}$, $H$, $\dot{H}$
all remain finite whereas $\ddot{H}$ becomes divergent. In this
universe the Bel Robinson energy at an initial time,
$\mathcal{B}(0)$, is finite whereas \textit{its time derivative}
is
\be\dot{\mathcal{B}}(\tau)=3\left[2\frac{\ddot{a}}{a}(\ddot{H}+2H\dot{H})+
4\left(\frac{k}{a^{2}}+H^{2}\right) \left(-\frac{k
H}{a^{2}}+H\dot{H}\right)\right]\ee and thus
$\dot{\mathcal{B}}(\tau)\rightarrow\infty$ at $\tau\rightarrow 0$.
This is like a $B_4$ singularity in our scheme. Since the
derivative of the Bel-Robinson energy diverges, we may interpret
this singularity geometrically as one \emph{in the velocities of
the Bianchi (frame) field.} At $t_s$ the Bianchi field encounters
a cusp and its velocity diverges there.

The character of the singularities can be expressed in terms of
the functions describing the matter fields. We shall end this
Section with a result of this sort. Note that the necessary
conditions for singularities given in \cite{ck05} and based on the
behaviour of the Hubble parameter alone  can be rephrased in terms
of the electric parts of Bel Robinson energy. We have the
following result concerning the nature of the spacetime
singularities in such models.
\begin{theorem}
Necessary conditions for null and timelike geodesically
incomplete, globally and regularly hyperbolic \textsc{FRW}
universes are that at a finite time:
\begin{description}
\item [$B_{1}$] $|E|\rightarrow \infty$ and $|D|\rightarrow
\infty$  $\Leftrightarrow$ $\mu\rightarrow \infty$ and
$|\mu+3p|\rightarrow\infty$

\item [$B_{2}$] $|E|<\infty$ and $|D|\rightarrow
\infty$ $\Leftrightarrow$ $\mu\rightarrow \infty$ and
$|\mu+3p|<\infty$

\item [$B_{3}$] $|E|\rightarrow \infty$ and $|D|<
\infty$ $\Leftrightarrow$ $\mu<\infty$ and
$|\mu+3p|\rightarrow\infty$ $\Leftrightarrow$ $\mu<\infty$ and
$|p|\rightarrow\infty$

\item [$B_{4}$] $|E|< \infty$ and $|D|<
\infty$ $\Leftrightarrow$ $\mu< \infty$ and $|\mu+3p|<\infty$
$\Leftrightarrow$ $\mu<\infty$ and $|p|<\infty$.
\end{description}
\end{theorem}
%%%%%%%%%%%%%%%%%%%%%%%%%%%%%%%%%%%%%%%%%%%%%%%%%

\section {Incompleteness and Bel-Robinson energy}

%%%%%%%%%%%%%%%%%%%%%%%%%%%%%%%%%%%%%%%%%%%%%%%%%

It was recently shown by Ellis in \cite{ellis} that a \textsc{RW}
space with scale factor $a(t)$ admits a past closed trapped
surface if the following condition is satisfied: \be \label{cts}
\dot{a}(t)>\left|\frac{f'(r)}{f(r)}\right|, \ee with $f(r)=\sin
r,r,\sinh r$ for $k=1,0,-1$ respectively. Recall that a closed
trapped surface is a 2-surface with spherical topology such that
both families of incoming and outgoing null geodesics orthogonal
to the surface converge. Since our function $|D|$ can be written
in the form \be |D|=\frac{\sqrt{3}}{a^{2}(t)}
\left|\left(\dot{a}(t)-\frac{f'(r)}{f(r)}\right)
\left(\dot{a}(t)+\frac{f'(r)}{f(r)}\right)+
\frac{1}{f^{2}(r)}\right|, \ee we see that the condition for the
existence of a closed trapped surface becomes equivalent to the
following inequality: \be \label{d}
|D|>\frac{\sqrt{3}}{a^{2}(t)f^{2}(r)}. \ee We thus conclude that
collapse singularities (as predicted by the existence of a trapped
surface) are characterized by a divergent Bel Robinson energy.

Consider for example the flat graduated inflationary model of
\cite{ba}. This universe satisfies the null energy condition
$\mu+p=\gamma \mu^{3/4}\geq 0$, $\gamma>0$ and the scale factor is
given by \be a(t)=\exp\left(-\frac{16}{3^{3/2}\gamma^{2}t}\right),
\ee so that the inequality \be
\dot{a}(t)=\frac{16}{3^{3/2}\gamma^{2}}\frac{1}{t^{2}}
\exp\left(-\frac{16}{3^{3/2}\gamma^{2}t}\right)>\frac{1}{r} \ee is
satisfied for large $r$. We  conclude that there exist past closed
trapped surfaces in this model and by the singularity theorems we
find that this model must be null geodesically incomplete. The
nature of this singular behaviour is described by saying that it
is of type $(S_{1},N_{1},B_{1})$.

Another example is the flat \textsc{RW} filled with a massless
scalar field given in \cite{fo}. This  satisfies the null energy
condition and has a past closed trapped surface:  for large $r$ we
have \be \dot{a}(t)=\frac{1}{3t^{2/3}}>\frac{1}{r}. \ee It is
therefore null geodesically incomplete  of type
$(S_{1},N_{1},B_{1})$.

As a last example, consider the sudden singularity first
introduced in \cite{ba04} which can arise in an \textsc{RW} model
with scale factor given by \be
a(t)=\left(\frac{t}{t_{s}}\right)^{q}(a_{s}-1)+1-
\left(1-\frac{t}{t_{s}}\right)^{n}, \ee where $1<n<2$ and $0<q\leq
1$. We see that at $t_{s}$, $a(t)\rightarrow a_{s}$,
$\dot{a}(t)\rightarrow \dot{a}_{s}$ and $\ddot{a}(t)\rightarrow
-\infty$. When $\mu\geq 0$ and $p\geq 0$, we find that there
exists a future closed trapped surface (the analogous inequality
of (\ref{cts}) is $\dot{a}(t)<-|f'(r)/f(r)|$) since we have that
the condition \be
\dot{a}(t)=\frac{q}{t_{s}^{q}}{t^{q-1}}(a_{s}-1)+
\frac{n}{t_{s}^{n}}(t_{s}-t)^{n-1}<-|\cot r| \ee is satisfied for
$r=\pi/2$, $a_{s}<1$ and $0<t<t_{s}$. Therefore this model is
timelike and null geodesically incomplete in the future with
$|E|\rightarrow \infty$ but $|D|<\infty$ (type
$(S_{3},N_{2},B_{3})$).

\ack This work was supported by the joint Greek Ministry of
Education and European Union research grants `Pythagoras' and
`Heracleitus' and this support is gratefully acknowledged.

\end{document}